\documentclass[conference]{IEEEtran}
\IEEEoverridecommandlockouts
\usepackage{cite}
\usepackage{amsmath,amssymb,amsfonts}
\usepackage{algorithm}
\usepackage{algcompatible}
\usepackage{graphicx}
\usepackage{bm}
\usepackage{textcomp}
\usepackage{hyperref}
\usepackage{xcolor}
\def\BibTeX{{\rm B\kern-.05em{\sc i\kern-.025em b}\kern-.08em
    T\kern-.1667em\lower.7ex\hbox{E}\kern-.125emX}}
    
\begin{document}

\title{Joint Activity Detection and Data Decoding in Massive Random Access via a Turbo Receiver\\
\thanks{This work was supported in part by a start-up fund of The Hong Kong Polytechnic University (Project ID P0038174) and in part by the General Research Fund (Project No. 15207220) from the Hong Kong Research Grants Council.}
}

\author{\IEEEauthorblockN{Xinyu Bian, Yuyi Mao, and Jun Zhang}
\IEEEauthorblockA{{Department of Electronic and Information Engineering} \\
{The Hong Kong Polytechnic University}, Hong Kong \\
Email: xinyu.bian@connect.polyu.hk, yuyi-eie.mao@polyu.edu.hk, jun-eie.zhang@polyu.edu.hk}
}

\maketitle

\begin{abstract}
In this paper, we propose a turbo receiver for joint activity detection and data decoding in grant-free massive random access, which iterates between a detector and a belief propagation (BP)-based channel decoder. Specifically, responsible for user activity detection, channel estimation, and soft data symbol detection, the detector is developed based on a bilinear inference problem that exploits the common sparsity pattern in the received pilot and data signals. The bilinear generalized approximate message passing (BiG-AMP) algorithm is adopted to solve the problem using probabilities of the transmitted data symbols estimated by the channel decoder as prior knowledge. In addition, extrinsic information is derived from the detector to improve the channel decoding accuracy of the decoder. Simulation results show significant improvements achieved by the proposed turbo receiver compared with conventional designs.
\end{abstract}

\begin{IEEEkeywords}
Grant-free massive random access, turbo receiver, approximate message passing (AMP), activity detection, channel coding.
\end{IEEEkeywords}

\section{Introduction}
Massive machine-type communications (mMTC) has been identified as one of the main use scenarios in the fifth generation (5G) wireless networks in order to accommodate massive Internet-of-Things (IoT) connections. In 5G mMTC, uplink traffic dominates, which necessitates efficient random access (RA) schemes that empower a base station (BS) to serve a large pool of users with sporadic activity patterns. Grant-based RA schemes, which are typically adopted in the long-term evolution (LTE) uplink, may incur long latency and significant signalling overhead, making it inapplicable to massive RA \cite{xchen2021}.

Grant-free RA, where each user directly transmits pilot and data without the need of acquiring access permission from the BS, becomes a promising solution for mMTC \cite{xchen2021}. As only a small proportion of the users dynamically become active, it is critical for the BS to perform accurate user activity detection based on the received pilot signal prior for data reception in grant-free RA. Nevertheless, identifying the set of active users from numerous candidates is by no means trivial, as assigning orthogonal pilot sequences to all the users is impossible with the limited radio resources for pilot transmissions.

Fortunately, since the sporadic user activity pattern in mMTC is inherently embedded in the received pilot signal at the BS, compressive sensing was shown to be effective for user activity detection \cite{shag2018}, which was also jointly investigated with data reception for grant-free massive RA. In \cite{bwang2016}, an orthogonal matching pursuit (OMP) based algorithm was proposed for joint activity and data detection in non-orthogonal multiple access (NOMA). A similar problem was investigated in \cite{cwei2017} via an alternative approach that fuses approximate message passing (AMP) with expectation maximization (EM). Nevertheless, these works assumed perfect channel state information (CSI) at the BS, which is idealized and motivates the joint estimation of user activity and channel knowledge. In \cite{lliuwyu2018}, a vector AMP algorithm was proposed for joint user activity detection and channel estimation in massive multiple-input multiple-output (MIMO) systems with massive connectivity. To improve the activity detection and channel estimation performance, wireless channel sparsities in both the spatial and angular domains were exploited in \cite{mke2020}.

While extensive attempts were made to fully utilize the received pilot signal, a key property of grant-free RA was mostly ignored, i.e., both the transmitted pilot and data symbols are distorted by the same wireless fading channel. In other words, the received data signal shares the common sparsity pattern with the received pilot, which facilitated the recent development of data-assisted approaches for grant-free massive RA. Specifically, a joint user activity, channel and data estimation algorithm based on bilinear generalized approximate message passing (BiG-AMP) and loop belief propagation (BP) was proposed in \cite{qzou2020}, where the estimated payload data is used to assist channel estimation. However, this study was limited to uncoded transmissions, while more information can be exploited from the channel decoder to further enhance the performance \cite{xbian2021}. This calls for a holistic consideration on the common sparsity pattern and soft information from a channel decoder for receiver designs in grant-free massive RA.

In this paper, we propose a turbo receiver for joint activity detection and data decoding in grant-free massive RA. Specifically, the turbo receiver iterates between a detector and a channel decoder. The detector, which is responsible for user activity detection, channel estimation, and soft data symbol detection, is developed based on BiG-AMP in order to exploit the common sparsity pattern in the received pilot and data signals. The estimated soft data symbols are then converted to the prior information on the coded bits, which is used for channel decoding. The channel decoder is designed based on BP, which calculates the log-likelihood ratios (LLRs) as soft decoding results. We use the LLRs to refine the prior information of the transmitted data symbols for more accurate activity detection, channel and data symbol estimation in the next iteration. Simulation results show that significant performance improvements are achieved by the proposed turbo receiver, in terms of user activity detection, channel estimation, and data decoding compared with two baseline schemes. Besides, it behaves close to a performance upper bound with perfect knowledge of the user activity pattern.

\textbf{Notations:} We use lower-case letters, bold-face lower-case letters, bold-face upper-case letters, and math calligraphy letters to denote scalars, vectors, matrices, and sets, respectively. The entry in the $i$-th row and $j$-th column of a matrix $\mathbf{M}$ is denoted as $m_{ij}$. Besides, we denote the exponential function as $\text{exp}\left(\cdot\right)$ and denote the complex Gaussian distribution with mean $\bm{\mu}$ and covariance matrix $\bm{\Sigma}$ as $\mathcal{C} \mathcal{N}(\bm{\mu}, \bm{\Sigma})$.

\section{System Model}
We consider an uplink cellular system with an $M$-antenna BS serving $N$ single-antenna users. At each transmission block, $K$ ($K\leq N$) among the $N$ users become active for transmission. We define $u_{n}\in\{0,1\}$ as the user activity indicator, where $u_{n}=1$ means user $n$ is active and $u_{n} = 0$ if it is inactive. The sets of system users and active users are denoted as $\mathcal{N}\triangleq \{1,\cdots, N\}$ and $\Xi \triangleq \left\{j \in \mathcal{N} | u_{j}=1 \right\}$, respectively, and the set of BS antennas is denoted as $\mathcal{M}\triangleq\{1,\cdots,M\}$. Each transmission block spans $T$ symbol intervals, and quasi-static block fading is assumed. The uplink channel vector from user $n$ to the BS is denoted as $\mathbf{f}_{n} \triangleq \sqrt{\beta_{n}} \bm{\alpha}_{n}$, where $\bm{\alpha}_{n}$ and $\beta_{n}$ correspond to the small-scale and large-scale fading coefficients, respectively. We assume $\{\beta_{n}\}$'s are constants known by the BS \cite{zchen2018} and $\bm{\alpha}_{n}\sim\mathcal{CN}\left(\mathbf{0},\mathbf{I}\right)$.

We adopt a two-phase grant-free RA scheme for uplink transmission, where in each transmission block, the first $L$ symbols, denoted as $\mathbf{T}_{p}$, are reserved for pilot transmission and the remaining $L_{d}\triangleq T-L$ symbols, denoted as $\mathbf{T}_{d}$, are used for data transmission. As $N\gg L$ in mMTC \cite{lliuwyu2018}, we assign the users with a set of non-orthogonal and unique pilot sequences $\sqrt{L}\mathbf{x}_{p n}$, where $\mathbf{x}_{p n}\triangleq \left[x_{n 1}, \cdots, x_{n L}\right] \sim \mathcal{C} \mathcal{N}\left(\bm{0},\mathbf{I}\right)$, and define $\mathbf{X}_{p}\triangleq \sqrt{L}\left[\mathbf{x}_{p 1}, \cdots, \mathbf{x}_{p N}\right]^{\mathrm{T}}$.

Each active user has $N_{b}$ payload bits, denoted as $\bm{b}_{n}\triangleq \left[b_{n 1}, \cdots, b_{n N_{b}}\right], n \in \Xi$, which are encoded before transmission. In particular, a code block $\bm{d}_{n} \triangleq \left[d_{n 1}, \cdots, d_{n N_{d}}\right]$ is formed by attaching the cyclic redundancy check (CRC) bits to the payload bits, which is then encoded by a channel code. The channel-coded bits are expressed as $\bm{c}_{n} \triangleq \left[c_{n 1}, \cdots, c_{n N_{c}}\right]$ and they are modulated to a set of constellation points $\mathcal{X}$ with normalized average power via an invertible mapping $\mu: \{0,1\}^{\log_{2}|\mathcal{X}|}\rightarrow \mathcal{X}$. We denote the modulated data symbols for user $n$ as $\mathbf{x}_{d n}\triangleq \left[x_{n (L+1)}, \cdots, x_{n T}\right]$. For simplicity, we assume $N_{c}=L_{d} \log_2|\mathcal{X}|$ and set \{$\mathbf{x}_{dn}$\}'s to zero vectors for the inactive users. Let $\mathbf{X}_{d}\triangleq \left[\mathbf{x}_{d 1}, \cdots, \mathbf{x}_{d N}\right]^{\mathrm{T}}$ be the transmitted data symbols from all users and define $\mathbf{X}\triangleq \left[\mathbf{X}_{p}, \mathbf{X}_{d}\right]$. The received signal at the BS can thus be expressed as follows: 
\begin{align}
\mathbf{Y}&=\sqrt{\gamma}\mathbf{H}\mathbf{X}+\mathbf{N},
\end{align}

\noindent where $\gamma$ is the uplink transmit power, $\mathbf{H} \triangleq \left[\mathbf{h}_{1},...,\mathbf{h}_{N}\right]$ with $\mathbf{h}_{n}\triangleq u_{n}\mathbf{f}_{n}$ denotes the effective channel coefficient matrix, and $\mathbf{N} \triangleq \left[\mathbf{n}_{1},...,\mathbf{n}_{T}\right]$ is the additive white Gaussian noise (AWGN) with zero mean and variance $\sigma^{2}$ for each element. Note that each element in $\mathbf{H}$ follows a Bernoulli-Gaussian distribution due to the sporadic user activity pattern.

In the next section, we will develop a turbo receiver for grant-free massive RA, where the common sparsity pattern of the received pilot and data signals as well as the soft decoding information are utilized.

\section{Turbo Receiver for Joint Activity Detection and Data Decoding}
\subsection{Overview}
\begin{figure}[htpb]
\centering
\includegraphics[width=3.4in]{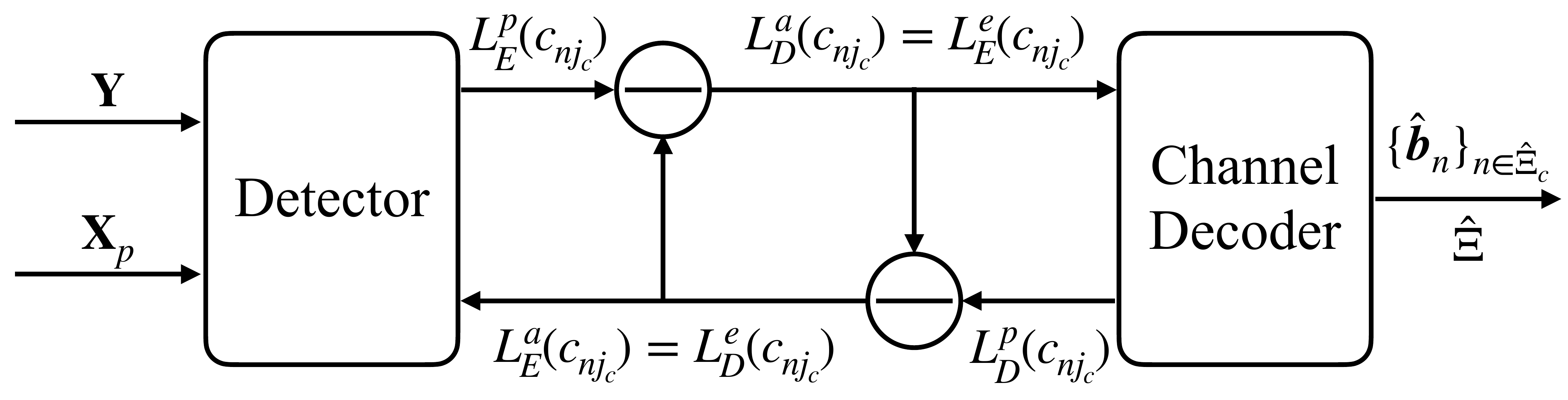}
\caption{The proposed turbo receiver for joint activity detection and data decoding in grant-free massive RA.}
\label{jointmodel}
\end{figure}

The proposed turbo receiver is motivated by the turbo decoder from coding theory \cite{bvu2001}, which iterates between a detector and a channel coder as shown in Fig. \ref{jointmodel}. The detector is responsible for user activity detection, channel estimation, and soft data symbol detection, which is developed based on BiG-AMP. It calculates the posterior probabilities of the transmitted data symbols, which are converted as input of the channel decoder. The BP algorithm is adopted by the channel decoder. It outputs the posterior log likelihood ratios (LLRs) for each bit in the transmitted code block as the soft decoding results, which are translated as prior information of the data symbols for the detector in the next turbo iteration. The turbo receiver terminates either when an exit condition is met or after $Q_{1}$ rounds of iteration, after which, hard decoding is performed.
\vspace{-0.2cm}
\vspace{-0.6cm}
\subsection{The Detector} 
\subsubsection{Design Objective}
The detector estimates the user activity, channel coefficients, and soft data symbols from the received signal $\mathbf{Y}$. Since the user activity pattern is captured by $\mathbf{H}$, we adopt a two-step approach for activity detection and channel estimation by first estimating the effective channel coefficients, followed by a simple thresholding step to determine the set of active users \cite{mke2020}. In particular, minimum mean square error (MMSE) estimators are adopted for both the effective channel coefficients and soft data symbols, which are given as the following expressions:
\vspace{-0.2cm}
\begin{align}
\forall m, n: \hat{h}_{mn}\triangleq\mathbb{E}\left[h_{mn}|\mathbf{Y}\right]=\int h_{mn} p(h_{mn}|\mathbf{Y}) dh_{mn},
\end{align}
\vspace{-0.4cm}
\begin{align}
\forall n, t \in \mathbf{T}_{d}: \hat{x}_{nt}\triangleq\mathbb{E}\left[x_{nt}|\mathbf{Y}\right]=\sum x_{nt} p(x_{nt}|\mathbf{Y}),
\end{align}

\noindent where $p(h_{mn}|\mathbf{Y})$ and $p(x_{nt}|\mathbf{Y})$ denote the marginal posterior distributions. They can be expressed in terms of the joint posterior distribution $p(\mathbf{H},\mathbf{X}|\mathbf{Y})$ as follows:
\vspace{-0.2cm}
\begin{align}
p\left(h_{mn}|\mathbf{Y}\right)=\int\nolimits_{\mathbf{H}_{\backslash m, n}} \sum\nolimits_{\mathbf{X}} p(\mathbf{H},\mathbf{X}|\mathbf{Y})d\mathbf{H},
\end{align}
\vspace{-0.2cm}
\begin{align}
p\left(x_{nt}|\mathbf{Y}\right)=\sum\nolimits_{\mathbf{X}_{\backslash n, t}} \int\nolimits_{\mathbf{H}} p(\mathbf{H},\mathbf{X}|\mathbf{Y})d\mathbf{H},
\end{align}

\noindent where $\mathbf{A}_{\backslash i, j}$ denotes all the elements in matrix $\mathbf{A}\triangleq[a_{ij}]$ expect $a_{ij}$. By using the Bayes' rule, $p(\mathbf{H},\mathbf{X}|\mathbf{Y})$ can be factorized as follows:
\vspace{-0.1cm}
\begin{align} \notag
    p(\mathbf{H},\mathbf{X}|\mathbf{Y})&=\frac{1}{Q} p(\mathbf{Y}|\mathbf{H}, \mathbf{X}) p(\mathbf{H|U})p(\mathbf{U}) p(\mathbf{X}_{p})p(\mathbf{X}_{d})\\ \notag
&=\frac{1}{Q} \prod\nolimits_{m=1}\nolimits^{M}\prod\nolimits_{t=1}\nolimits^{T} p\left(y_{m t}|\sum\nolimits_{n=1}\nolimits^{N} h_{m n} x_{n t}\right) \notag
\end{align}
\vspace{-0.2cm}
\begin{align}
\times \prod\nolimits_{n=1}\nolimits^{N}\left[p\left(u_{n}\right) \prod\nolimits_{m=1}\nolimits^{M} p\left(h_{m n}|u_{n}\right) \prod\nolimits_{t=1}\nolimits^{T} p\left(x_{n t}\right)\right],
\end{align}

\noindent where $Q$ is a normalizing constant.

Nevertheless, computing the marginal distributions in (4) and (5) is intractable due to the high-dimensional integrals and summations. While exact computation of these quantities are prohibitive, inspired by the factorization in (6), we resort to the framework of AMP to derive computation-efficient approximations with near MMSE performance. Specifically, according to the signal model in (1), the problem of effective channel coefficient estimation and soft data symbol detection can be considered as a \emph{generalized bilinear inference} problem \cite{jtparker2015}. To solve the problem, we resort to the BiG-AMP algorithm, which provides a viable solution with affordable complexity. We note that different from traditional blind signal detection algorithms, the received pilot symbols are jointly utilized by the detector.

\subsubsection{Algorithm Design}
The proposed detector based on BiG-AMP is summarized in Algorithm 1. For brevity, we omit the iteration number. There are four main steps in Algorithm 1: First, the linear mixing variable $z_{m t}\triangleq\sum_{n=1}^{N} h_{m n} x_{n t}$ is estimated based on the received signal $\mathbf{Y}$. The basic principle is that if the prior information of a variable and its likelihood function under some observations are available, the posterior probability can be calculated using the Bayes' rule. With the posterior probability $p(z_{mt}|y_{mt})$, $z_{mt}$ can then be estimated by an MMSE estimator. We approximate $z_{m t}$ as a complex Gaussian distribution, and its mean $M_{m t}^{p}$ and variance $V_{m t}^{p}$ are calculated in Line 3 and 4 in Algorithm 1 respectively according to BP. Since $\mathbf{Y}=\sqrt{\gamma}\mathbf{Z}+\mathbf{N}$, $p(y_{mt}|z_{mt})$ is a complex Gaussian distribution. Thus, the posterior distribution of $z_{mt}$ is also complex Gaussian with its mean and variance computed in Line 5 and 6, respectively. Note that for pilot symbols, we have the exact estimate, i.e., $\hat{x}_{nt}={x}_{nt}$, and the variance is thus given by $V_{nt}^{x}=0$. Second, the scaled residual $\hat{s}_{mt}$ of the linear mixing variable $z_{mt}$ and the corresponding inverse-residual-variance $V_{m t}^{s}$ are calculated in Line 7 and 8, which are used to approximate the likelihood functions of the effective channel coefficients and data symbols in the following two steps respectively. In the third step, the effective channel coefficients and their variances are estimated with an MMSE estimator by incorporating both the pilot and data symbols in Line 18 and 19, respectively, where $P_{mn}^{h}$ ($Q_{mn}^{h}$) are the conditional mean (variance) of $h_{mn}$. Key procedures to obtain these values are summarized in Lines 9-14. Observing that $h_{mn}$ follows a Bernoulli-Gaussian distribution, we calculate the \emph{sparsity level} $\rho_{mn}$, i.e., the probability that $h_{mn}$ is non-zero, based on BP in Line 15 and 16, and the posterior sparsity level $\tilde{\rho}_{mn}$ is derived in Line 17 accordingly. Note that estimates of the likelihood that each user is active in the considered transmission block, i.e., $\{\lambda_{n}\}$'s, are required in Line 15. However, as such prior knowledge is unknown, we simply set $\lambda_{n}=K/N$, $\forall n \in \mathcal{N}$. Following similar procedures as the estimation of effective channel coefficients, the soft data symbols $\hat{x}_{nt}$ and the probabilities that a data symbol belongs to the constellation point $s$, denoted as $\{\tilde{\eta}_{nt,s}\}$'s, are estimated in the last step (Lines 20-24).

When Algorithm 1 terminates, i.e., the normalized difference of $\hat{z}_{mt}$ between neighboring iterations is negligible or the maximum number of iterations is reached, we determine the set of active users as $\hat{\Xi}\triangleq \{n\in\mathcal{N}|\frac{1}{M}\sum_{m=1}^{M} \tilde{\rho}_{mn} \geq\theta\}$, where $\theta$ is an empirical threshold \cite{mke2020}.

\subsubsection{Extrinsic Information for the Decoder}
With the estimated set of active users and the soft data symbols, we derive the extrinsic information of the detector that serves as input of the channel decoder. The objective is to remove some redundancy from the prior information of the coded bits so that the channel decoding accuracy can be improved. To achieve this, the posterior probabilities of each data symbol are first converted to the posterior probabilities of the corresponding coded bits:
\vspace{-0.3cm}
\begin{align}
p(c_{nj_{c}}=b|\mathbf{Y})=\sum \nolimits_{s \in \mathcal{X}_{\hat{j}_{c}}^{b}}\tilde{\eta}_{nt,s}, n \in \hat{\Xi},
\end{align}
\noindent where $\hat{j}_{c}\triangleq \mod({j}_c,\log_{2}|\mathcal{X}|)$, $t=L+1+\lfloor \frac{j_{c}}{\log_{2}|\mathcal{X}|} \rfloor$, and $\mathcal{X}_{l}^{b}$ represents the set of constellation points with the $l$-th position ($l=0,\cdots, \log_{2}|\mathcal{X}|-1$) of the corresponding bit sequence as $b$. For example, suppose the bit sequences ‘00’, ‘01’, ‘10’ and ‘11’ are modulated to $s_{0}$, $s_{1}$, $s_{2}$, and $s_{3}$ respectively in quadrature phase shift keying (QPSK), we have $\mathcal{X}_{0}^{0} = \{s_{0},s_{1}\}$, $\mathcal{X}_{0}^{1} = \{s_{2},s_{3}\}$, $\mathcal{X}_{1}^{0} = \{s_{0},s_{2}\}$, and $\mathcal{X}_{1}^{1} = \{s_{1},s_{3}\}$. The posterior LLR of each coded bit is derived as follows:
\vspace{-0.3cm}
\begin{align}
L_{E}^{p}(c_{nj_{c}})\triangleq \ln \left(\frac{p(c_{nj_{c}}=0|\mathbf{Y})}{p(c_{nj_{c}}=1|\mathbf{Y})}\right), n \in \hat{\Xi}.
\end{align}

The extrinsic information of the detector is defined as the difference between the posterior and prior LLR of each coded bit as $L_{E}^{e}(c_{nj_{c}})\triangleq L_{E}^{p}(c_{nj_{c}})-L_{E}^{a}(c_{nj_{c}}), n \in \hat{\Xi}$, where $L_{E}^{a}(c_{nj_{c}})\triangleq \ln \left(\frac{p(c_{nj_{c}}=0)}{p(c_{nj_{c}}=1)}\right)$ denotes the prior LLR obtained from the decoder in the previous turbo iteration.
\addtolength{\topmargin}{0.01in}
\begin{algorithm}[t]
\caption{The Detector based on BiG-AMP}
{\bf Input:}
Received signal $\mathbf{Y}$, pilot symbols $\mathbf{X}_{p}$\\
{\bf Output:}
The estimated set of active users $\hat{\Xi}$, and the posterior probabilities that $x_{nt}$ equals $s$, i.e., $\tilde{\eta}_{nt,s}$, $n \in \hat{\Xi}$, $t \in \mathbf{T}_{d}$.\\
{\bf Initialize:}
$i \leftarrow 0$, $\hat{h}_{mn}(0)$, $V_{mn}^{h}(0)$, $s_{mt}(0)$, $\hat{x}_{nt}(0)$, $V_{nt}^{x}(0)$

\begin{algorithmic}[1]
\WHILE{{$i \leq Q_{2}$} \textbf{and} {$\frac{\Sigma_{m,t}|\hat{z}_{m t}(i)-\hat{z}_{m t}(i-1)|^{2}}{\Sigma_{m,t}|\hat{z}_{m t}(i-1)|^{2}}>\epsilon_{1}$}}
\STATE $i \leftarrow i+1$
\Statex // \textit{Estimate the linear mixing variable ${z}_{m t}$} //

\STATE $\forall m,t: M_{m t}^{p}(i)=\sum_{n} \hat{h}_{m n}(i-1) \hat{x}_{n t}(i-1)-\hat{s}_{m t}(i-1)\sum_{n}(|\hat{x}_{n t}(i-1)|^{2} V_{m n}^{h}(i-1)+|\hat{h}_{m n}(i-1)|^{2} V_{n t}^{x}(i-1))$

\STATE $\forall m,t: V_{m t}^{p}(i)=\sum_{n}(|\hat{x}_{n t}(i-1)|^{2} V_{m n}^{h}(i-1)+|\hat{h}_{m n}(i-1)|^{2} V_{n t}^{x}(i-1))+\sum_{n} V_{m n}^{h}(i-1)V_{n t}^{x}(i-1)$

\STATE $\forall m,t: \hat{z}_{m t}(i)=\mathbb{E}[z_{mt}|M_{m t}^{p}(i),V_{m t}^{p}(i)]$

\STATE $\forall m, t:V_{m t}^{z}(i)=\text{Var}[z_{mt}|M_{m t}^{p}(i),V_{m t}^{p}(i)]$

\Statex // \textit{Calculate the scaled residual and variance} //

\STATE $\forall m, t: \hat{s}_{m t}(i)=(\hat{z}_{m t}(i)-M_{m t}^{p}(i))/V_{m t}^{p}(i)$

\STATE $\forall m, t: V_{m t}^{s}(i)=(1-V_{m t}^{z}(i)/V_{m t}^{p}(i))/V_{m t}^{p}(i)$

\Statex // \textit{Estimate the effective channel coefficients} //

\STATE $\forall m, n: Q_{p,m n}^{h}(i)=\left(\sum_{t \in \mathbf{T}_{p}}|x_{nt}|^{2}V_{m t}^{s}(i)\right)^{-1}$

\STATE $\forall m, n: P_{p,m n}^{h}(i)=\hat{h}_{mn}(i-1)+Q_{p,m n}^{h}(i)$
$\sum_{t \in \mathbf{T}_{p}}x_{nt}^{*}\hat{s}_{m t}(i)$

\STATE $\forall m, n: Q_{d,m n}^{h}(i)=\left(\sum_{t \in \mathbf{T}_{d}}|\hat{x}_{nt}(i-1)|^{2}V_{m t}^{s}(i)\right)^{-1}$

\STATE $\forall m, n: P_{d,m n}^{h}(i)=\hat{h}_{mn}(i-1)(1-Q_{d,m n}^{h}(i)\sum_{t \in \mathbf{T}_{d}}$
$V_{n t}^{x}(i-1)V_{m t}^{s}(i))+Q_{d,m n}^{h(j)}(i)\sum_{t\in \mathbf{T}_{d}}\hat{x}_{nt}^{*}(i-1)\hat{s}_{m t}(i)$

\STATE $\forall m, n: Q_{m n}^{h}(i)=\frac{Q_{p,m n}^{h}(i)Q_{d,m n}^{h}(i)}{Q_{p,m n}^{h}(i)+Q_{d,m n}^{h}(i)}$

\STATE $\forall m, n: P_{m n}^{h}(i)=\frac{P_{p,m n}^{h}(i)Q_{d,m n}^{h}(i)+P_{d,m n}^{h}(i)Q_{p,m n}^{h}(i)}{Q_{p,m n}^{h}(i)+Q_{d,m n}^{h}(i)}$

\STATE $\forall m, n: L_{mn}(i)=\text{ln}\frac{\lambda_{n}}{1-\lambda_{n}}+\sum_{k \in \mathcal{M} \backslash \{m\}}(\text{ln}\frac{Q_{k n}^{h}(i)}{Q_{k n}^{h}(i)+\beta_{n}}+\frac{|P_{kn}^{h}(i)|^{2}\beta_{n}}{(Q_{k n}^{h}(i)+\beta_{n})Q_{k n}^{h}(i)})$

\STATE $\forall m, n: \rho_{mn}(i)=\frac{\text{exp}(L_{mn}(i))}{1+\text{exp}(L_{mn}(i))}$

\STATE $\forall m, n: \tilde{\rho}_{mn}(i)=\rho_{mn}(i)/(\rho_{mn}(i)+(1-\rho_{mn}(i))\text{exp}(-\text{ln}\frac{Q_{m n}^{h}(i)}{Q_{m n}^{h}(i)+\beta_{n}}-\frac{|P_{mn}^{h}(i)|^{2}\beta_{n}}{(Q_{m n}^{h}(i)+\beta_{n})Q_{m n}^{h}(i)}))$
\STATE $\forall m, n:\hat{h}_{mn}(i)=\mathbb{E}[h_{mn}|P_{m n}^{h}(i),Q_{m n}^{h}(i)]$

\STATE $\forall m, n: V_{m n}^{h}(i)=\text{Var}[h_{mn}|P_{m n}^{h}(i),Q_{m n}^{h}(i)]$

\Statex // \textit{Estimate the soft data symbols} //

\STATE $\forall n, t \in \mathbf{T}_{d}: Q_{n t}^{x}(i)=\left(\sum_{m}|\hat{h}_{mn}(i-1)|^{2}V_{m t}^{s}(i)\right)^{-1}$

\STATE $\forall n, t \in \mathbf{T}_{d}:P_{n t}^{x}(i)=\hat{x}_{nt}(i-1)(1-Q_{n t}^{x}(i)$
$\sum_{m}V_{m n}^{h}(i-1)V_{m t}^{s}(i))+Q_{n t}^{x}(i)\sum_{m}\hat{h}_{mn}^{*}(i-1)\hat{s}_{m t}(i)$

\STATE $\forall n, t \in \mathbf{T}_{d}:\tilde{\eta}_{nt,s}(i)=\frac{p(x_{nt}=s) \text{exp}(-|s-P_{nt}^{x}(i)|^{2}/Q_{n t}^{x}(i))}{\sum_{s'}p(x_{nt}=s')\text{exp}(-|s'-P_{nt}^{x}(i)|^{2}/Q_{n t}^{x}(i))}$

\STATE $\forall n, t \in \mathbf{T}_{d}:\hat{x}_{nt}(i)=\mathbb{E}[x_{nt}|P_{n t}^{x}(i),V_{n t}^{x}(i)]$

\STATE $\forall n, t \in \mathbf{T}_{d}: V_{n t}^{x}(i)=\text{Var}[x_{nt}|P_{n t}^{x}(i),V_{n t}^{x}(i)]$
\ENDWHILE
\end{algorithmic}
\end{algorithm}

\subsection{The Channel Decoder}
The channel decoder aims at finding the most probable code block for each active user based on the output of the detector, which can be mathematically written as the following maximum \emph{a posteriori} probability (MAP) estimation problem:
\begin{align}
\bm{\hat{c}}_{n}=\arg \max_{\bm{c}_{n} \in \{0,1\}^{N_{c}}} p(\bm{c}_{n} \mid \{\tilde{\eta}_{nt,s}\}), n \in \hat{\Xi}.
\end{align}

Considering its supreme effectiveness in calculating marginal distributions, BP has been widely used to design channel decoders. As a result, we adopt an off-the-shelf BP-based channel decoder to solve (9), which accepts prior information of the coded bits, i.e., $L_{E}^{e}(c_{nj_{c}})$, as input, denoted as $L_{D}^{a}(c_{nj_{c}})$, and operates on a Tanner graph of the code \cite{bvu2001} based on BP to calculate the posterior LLRs of the coded bits, as soft decoding information, denoted as $L_{D}^{p}(c_{nj_{c}})$. 

Similar to the detector, extrinsic information of the decoder can be calculated as $L_{D}^{e}(c_{nj_{c}}) \triangleq L_{D}^{p}(c_{nj_{c}})-L_{D}^{a}(c_{nj_{c}}), n\in \hat{\Xi}$, and prior distribution of each coded bit is derived as $p(c_{nj_{c}}=1)=\frac{1}{1+\text{exp}(L_{D}^{e}(c_{nj_{c}}))}$ and $p(c_{nj_{c}}=0)=\frac{\text{exp}(L_{D}^{e}(c_{nj_{c}}))}{1+\text{exp}(L_{D}^{e}(c_{nj_{c}}))}$, which are translated as prior information of the transmitted data symbols for the use of the detector in Line 22 of Algorithm 1 as follows:
\vspace{-0.2cm}
\begin{align}
p(x_{nt})=\prod \nolimits_{j_{c}=v_{1}}\nolimits^{v_{2}}p(c_{nj_{c}}), t \in \mathbf{T}_{d}, n\in \hat{\Xi}.
\end{align}

\noindent where $v_{1}\triangleq (t-L-1)\text{log}_{2}|\mathcal{X}|$, $v_{2}\triangleq (t-L)\text{log}_{2}|\mathcal{X}|-1$, and $\mu\left([c_{nv_{1}},\cdots,c_{nv_{2}}] \right) = x_{nt}$. Note that for the users that are determined as inactive, we reuse the prior information of the transmitted data symbols from the last turbo iteration.

After $Q_{1}$ rounds of turbo iteration, the code block $\hat{\bm{d}}_{n},n \in \hat{\Xi}$ is obtained from $\hat{\bm{c}}_{n},n \in \hat{\Xi}$ via hard decision as follows:
\vspace{-0.2cm}
\begin{align}
\hat{d}_{nj_{c}}=\left\{
\begin{aligned}
0,  L_{D}^{p}(c_{nj_{c}})\geq 0, \\
1,  L_{D}^{p}(c_{nj_{c}})<0.
\end{aligned}
\right.
\end{align}

\noindent If $\hat{\bm{d}}_{n}$ passes CRC, the payload bits are obtained by detaching the CRC bits, and the set of users that pass CRC is denoted as $\hat{\Xi}_{c}$. Otherwise, a decoding error occurs.

\section{Simulation Results}
We simulate an uplink cellular network with $200$ users randomly distributed in a circle with a radius of 500 m centered at the BS. The path loss is calculated as $\beta_{n}=-128.1-36.7\text{log}_{10}r_{n}$ (dB) for user $n$, where $r_{n}$ denotes its distance to the BS. The simulation results are averaged over $10^5$ independent channel realizations. Algorithm 1 is initialized based on AMP algorithm \cite{mke2020} and a damping factor is applied for robustness. Key simulation parameters are summarized in TABLE \ref{table1}.
\vspace{-0.5cm}
\begin{table}[ht]
\caption{Simulation parameters}
\centering
    \begin{tabular}{c|c|c|c}
    \hline
     Parameters & Values & Parameters & Values\\
    \hline
    $M$ & 64 & $T$ & 200 \\
    $N_{b}$ & 142 & $N_{d}$ & 150 \\
    $N_{c}$ & 300 & CRC type & CRC-8\\
    $L$ & 50 & $L_{d}$ & 150 \\
    $\theta$ & 0.4 & $Q_{1}$ & 3\\
    $Q_{2}$ & 100 & $\epsilon_{1}$ & $10^{-5}$ \\
    Bandwidth & 1 MHz &Modulation & QPSK\\
    Channel coding & LDPC & Code rate & $1/2$ \\
    Transmit power & 23 dBm & Noise power density & -169 dBm/Hz\\
    \hline
    \end{tabular}
\label{table1}
\end{table}

We also implement two baseline schemes and a performance upper bound in our simulations for comparisons:

\begin{itemize}
    \item \textbf{Separate design \cite{mke2020}:} This scheme first performs channel estimation and user activity detection using the AMP algorithm. Data symbols are then detected via an MMSE estimator, followed by a BP-based LDPC decoder.
\end{itemize}

\begin{itemize}
    \item \textbf{Data-assisted design with BiG-AMP:} This scheme uses the BiG-AMP algorithm for user activity detection, channel estimation and data symbol detection. The payload bits are decoded with a BP-based LDPC decoder. Note that this scheme is a special case of the proposed turbo receiver when $Q_{1}=1$.
\end{itemize}

\begin{itemize}
    \item \textbf{Turbo receiver with perfect activity pattern:} This scheme assumes perfect knowledge of the user activity pattern and performs channel estimation and data decoding via the proposed turbo receiver. It serves as a performance upper bound.
\end{itemize}

\begin{figure}[htpb]
\centering
\includegraphics[width=2.55in]{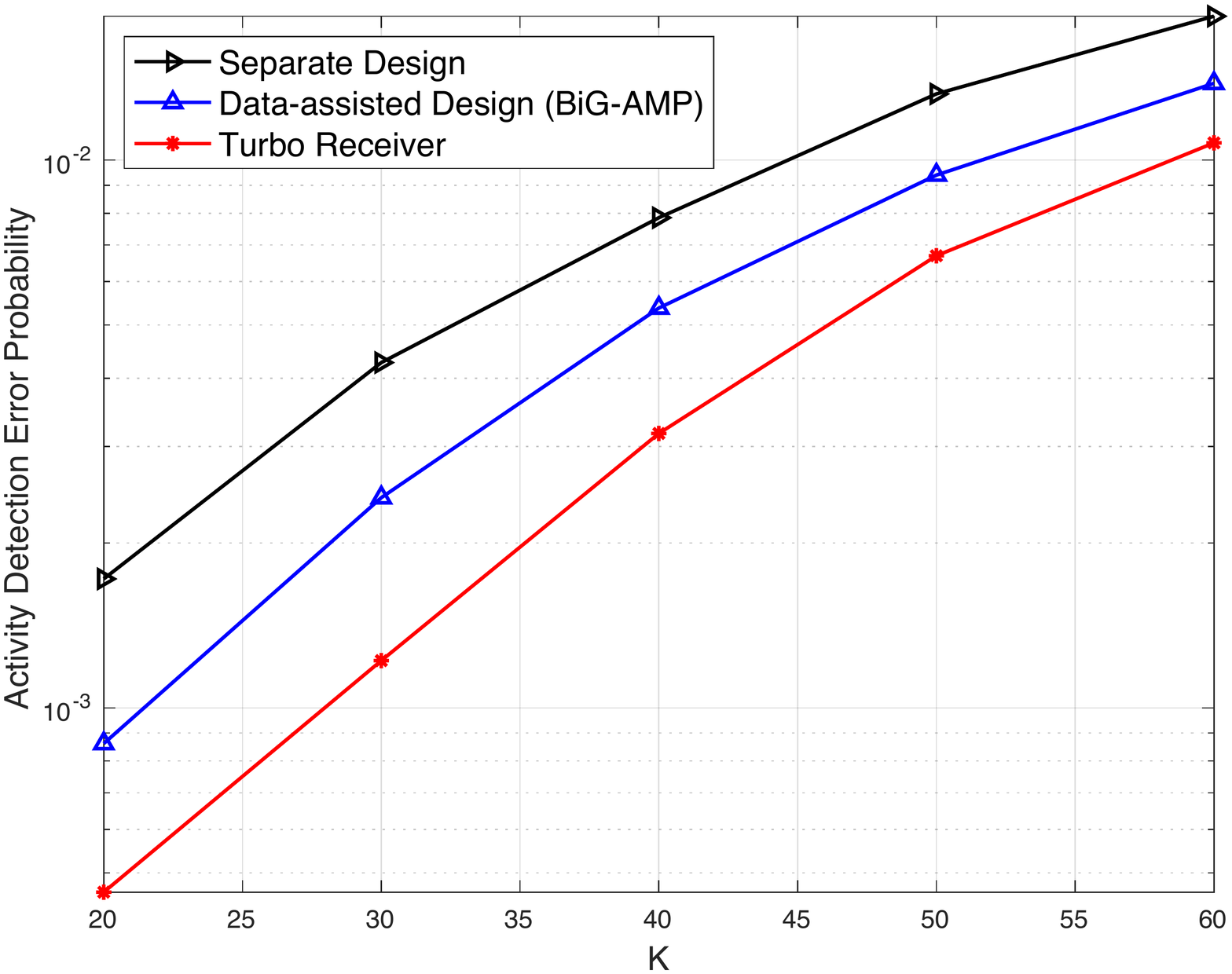}
\caption{Activity detection error probability vs. the number of active users.}
\label{S11}
\end{figure}

\begin{figure}[htpb]
\centering
\includegraphics[width=2.6in]{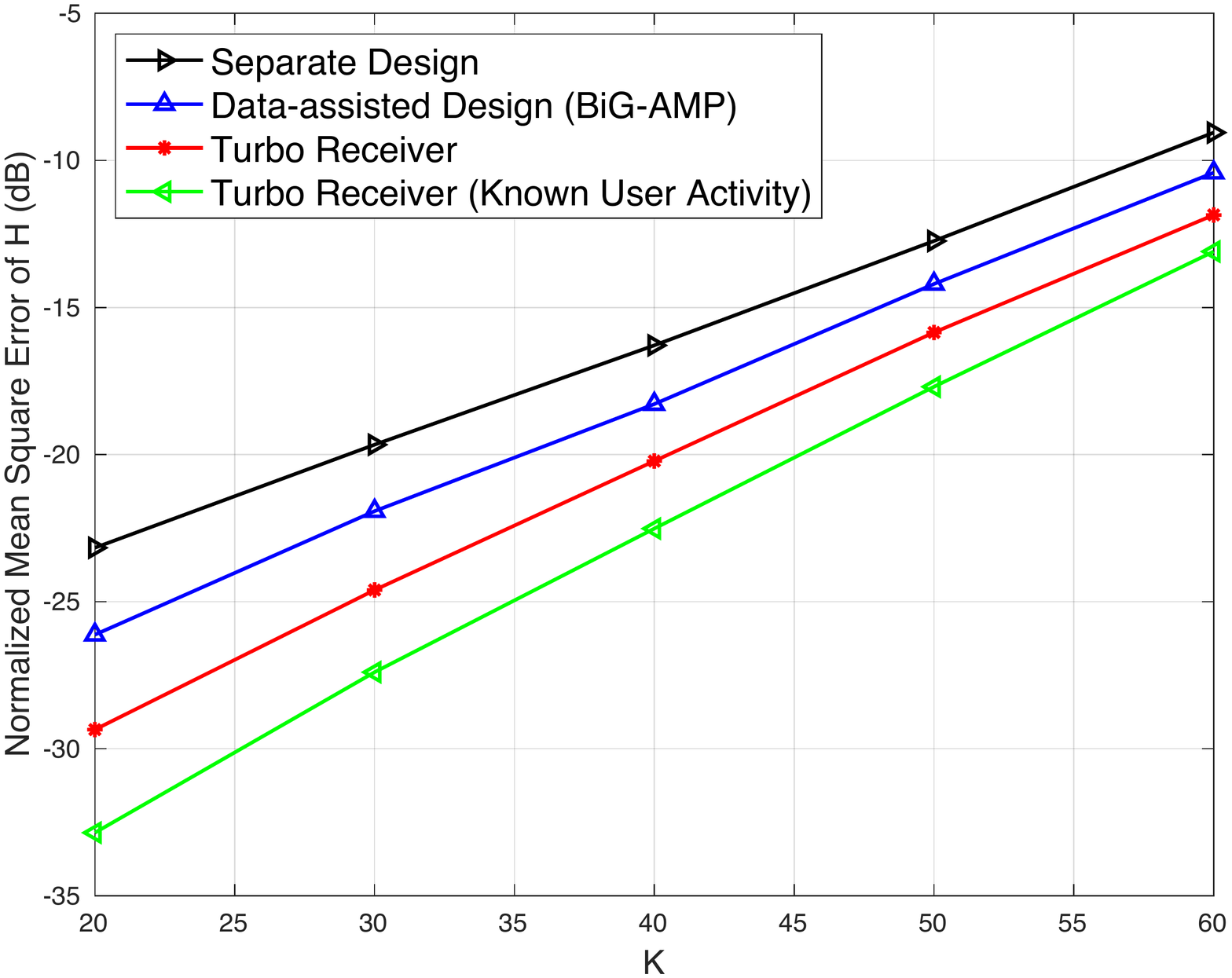}
\caption{NMSE of channel estimation vs. the number of active users.}
\label{S12}
\end{figure}

We first show the activity detection error probability and the normalized mean square error (NMSE) of channel estimation in Fig. \ref{S11} and Fig. \ref{S12}, respectively. From both figures, the performance degrades with the number of active users due to the limited radio resources for pilot transmissions. Compared to the separate design, the data-assisted design with BiG-AMP effectively reduces the user activity detection and channel estimation error as the common sparsity pattern in both the received pilot and data signals is exploited. Besides, the proposed turbo receiver secures noticeable further performance enhancements compared to the data-assisted design with BiG-AMP, which is attributed to the wise integration of the soft extrinsic information from the BP-based channel decoder that progressively refines prior information of data symbols.
\setlength{\abovecaptionskip}{-0.2cm} 
\vspace{-0.2cm}
\setlength{\abovecaptionskip}{-0.2cm} 
\vspace{-0.2cm}

\begin{figure}[htpb]
\centering
\includegraphics[width=2.55in]{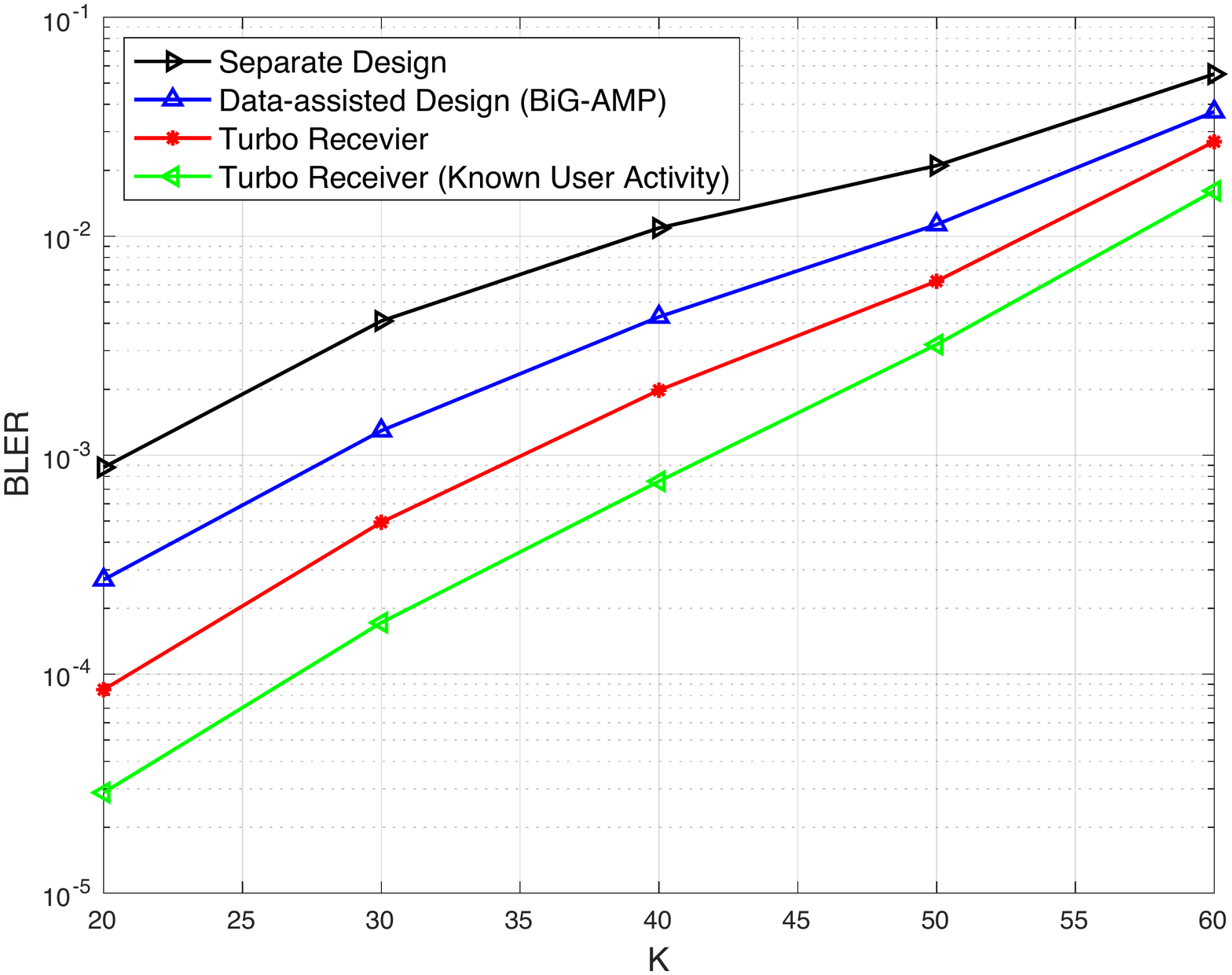}
\caption{BLER vs. the number of active users.}
\label{S2}
\end{figure}

The BLER performance of different schemes versus the number of active users is examined in Fig. \ref{S2}. In line with the user activity detection error and channel estimation performance, the proposed turbo receiver achieves significant BLER reduction compared to the separate design and data-assisted design with BiG-AMP. For instance, with a BLER requirement of $10^{-3}$, the turbo receiver is able to support about $75\%$ and $25\%$ more active users compared to the two baselines, respectively. It also behaves close to the case with perfect knowledge of user activity, which is even more remarkable with a large number of active users, e.g., $K \geq 50$. This again validates the benefits of exploiting the structural information of the received signal, as well as the soft decoding information in achieving reliable communications for massive RA.
\vspace{-0.2cm}
\vspace{-0.2cm}
\section{Conclusions}
In this paper, we proposed a turbo receiver for joint activity detection and data decoding for grant-free massive RA. On one hand, the common sparsity in the received pilot and data signals was exploited via BiG-AMP in the detector, whose output was used as prior information for a BP-based channel decoder. On the other hand, by removing redundancy of the decoder output, extrinsic information was used as the prior information in the detector for more accurate data symbol detection. Simulation results demonstrated that with the proposed framework, significant user activity detection, channel estimation, and data decoding errors can be reduced.

\end{document}